\documentclass
[journal,11pt,draftclsnofoot,onecolumn,doublespace]{tETN2e}

\usepackage{rotating}

\begin{document}
\title{{\itshape Collision Avoidance in TV White Spaces: %
A Cross-layer Design Approach for Cognitive Radio Networks %
} }

\author{Fotis Foukalas $^{a}$ $^{\ast}$\thanks{$^\ast $Corresponding %
author. Email: foukalas@di.uoa.gr \vspace{6pt}} and George T. %
Karetsos$^{b}$\\\vspace{6pt} $^{a}${\em{Dept. Informatics and  %
Telecommunications, University of Athens, Athens Greece}}; \\
$^{b}${\em{Dept. Informatics and Telecommunications, Technological %
Education Institute of Larissa, Larissa, Greece}}\\\vspace{6pt}%
\received{v1.0 released October 2012} }

\maketitle
\begin{abstract}
One of the most promising applications of cognitive radio networks (CRNs)is the efficient exploitation of TV white spaces (TVWSs) for enhancing the performance of wireless networks. In this paper, we propose a cross-layer design (CLD) of  carrier sense multiple access with collision avoidance (CSMA/CA) mechanism at the medium access control (MAC) layer with spectrum sensing (SpSe) at the physical layer, for identifying the occupancy status of TV bands. The proposed CLD relies on a Markov chain model with a state pair containing both the SpSe and the CSMA/CA from which we derive the collision probability and the achievable throughput. Analytical and simulation  results are obtained for different collision avoidance and spectrum sensing implementation scenarios  by varying the contention window, backoff stage and probability of detection. The obtained results depict the achievable throughput under different collision avoidance and spectrum sensing implementation scenarios indicating thereby the performance of collision avoidance in TVWSs based cognitive radio networks. 
\end{abstract}

   
\section{Introduction}
In the last decade, cognitive radio (CR) technology has been one of the main focal points by both the research community and the wireless communications industry. This is attributed to the fact that the CR technology promises an efficient and pervasive way for the utilization of spectrum resources. For instance, the Federal Communications Commission (FCC) in the US has identified that there are several unused portions in the TV spectrum range that are known as TV white spaces (TVWSs) \citep{fcc08}. However, CR technology requires specific functionalities in order to learn about the spectrum occupancy and then identify the quality of the TVWSs including the usage percentage from licensed users. Such a functionality can be provided by a spectrum sensing (SpSe) mechanism which can sense and provide feedback regarding spectrum occupancy \citep{yucek09} (Zeng et al. 2010). Besides, since the TVWSs should be available to different users at different locations the concept of cognitive radio networks (CRNs) has been defined \citep{akyildiz06,bixio11}. Such a network specifies all these functionalities making available the white spaces to cognitive users (CUs) through specific medium access control (MAC) techniques and protocols \citep{xiao08}. Due to its importance, in this paper, we focus on the collision avoidance functionality at the MAC layer of CRNs, where the latter exploits TVWSs.

In CRNs, channels’ availability is manifested via a SpSe mechanism at the physical (PHY) layer \citep{foukalas12} and subsequently the packet transmission is accomplished through an appropriate medium access control (MAC) layer technique \citep{cormio09, yan10}. In this case, an important issue at the MAC layer is the design of the collision avoidance mechanism when a CRN is implemented in a distributed coordination function (DCF). Looking into the current proposals for implementing the collision avoidance in CRNs, such a functionality is provided in two cases in IEEE 802.22 standard: a) when dynamic frequency hopping (DFH) between adjacent base stations (BSs) is supported, where both SpSe and data transmission takes place from multiple adjacent cells and b) when the point-to-point communication method (i.e. mesh type of CRN) is supported where the traffic flows among the customer premises equipments (CPEs) directly \citep{ieee06}. Moreover, the ECMA-392 standard provides a DCF through a prioritized carrier sense multiple access for collision avoidance (CSMA/CA) that permits multiple devices to contend for gaining access to the medium based on traffic priority \citep{ecma09}.

Nevertheless, the incorporation of the MAC layers with the PHY layer has not been considered so far in order to guarantee the efficient interworking between these two layers. In this paper, we investigate the collision avoidance in CRNs operating over TVWSs using a cross-layer design (CLD) approach. Using the proposed CLD, we incorporate the multi-channel SpSe at the PHY layer with the CSMA/CA protocol at the MAC layer which employs exponential backoff and saturation conditions. In the proposed CLD approach, both SpSe and CSMA/CA are modelled as a two-dimensional Markov chain. From the steady state convergence of the stochastic Markov chain analysis, we derive the collision probability that is used in the sequel for deriving the achievable throughput. Both collision probability and throughput are considered the performance metrics in this particular investigation. The obtained results illustrate the achievable throughput under different collision avoidance conditions and spectrum sensing scenarios by varying the contention window, backoff stage and probability of detection.  

To the best of our knowledge and based on the literature review presented below, such design has not been investigated so far by the research community. However, there exist a few solutions, which are relative to the proposed CLD that we present subsequently. In \citep{chen08} and \citep{liang08} a CLD between the SpSe mechanism at the physical layer and the MAC layer is proposed in which a constraint on collision avoidance dictates the operating characteristics of SpSe; however, neither a specific collision avoidance protocol nor any other protocol has been incorporated. In (Chong, 2010), the authors highlight the impact of imperfect SpSe at the physical layer on the performance of the MAC layer by using a cross-layer performance analysis; however, they have not provided a specific CLD scheme and most importantly their analysis does not take into account exponential backoff for the CSMA/CA protocol and multichannel SpSe. Therefore, a study that considers multi-channel SpSe, exponential backoff capabilities of collision avoidance in CRNs and a CLD with SpSe has not been provided so far and thereby this is the aim of this work. 

The rest of this paper is organized as follows. In section 2 the proposed cross-layer design is presented while in section 3 the performance metrics are derived through an appropriate analysis, which lead to the formulation of collision probability and achievable throughput expressions. Section 4 presents and discusses the results and we conclude with a brief discussion in section 5.

\section{Cross-layer Design of Collision Avoidance with Spectrum Sensing in CRNs}

We assume $n$ stations for the cognitive radio network, which are spatially distributed and are sharing a pool of channels. Each station is able to sense $c\in(1,C)$ channels using the spectrum sensing (SpSe) mechanism at the physical layer. SpSe is accomplished via an energy detection scheme over an additive white Gaussian noise (AWGN) channel or under Rayleigh fading. Energy detection is the most widely used technique since it has low implementation complexity for unknown signals \citep{yucek09}. The energy detection scheme senses the signal to noise ratio (SNR) $\gamma$ of the primary signal for a sensing time $T_s$ at frequency $f_s$. Considering imperfect SpSe, a channel is assumed idle when the SNR $\gamma$ is less than a predefined threshold value $\eta$, i.e. $\gamma < \eta$, and thus four possible cases are introduced for the state of the primary network (PN) as their perceived at the secondary network (SN) known as: detection, missed detection, false alarm and no false alarm with probabilities $P_d$ , $(1-P_d)$, $P_f$ and $(1-P_f)$ respectively \citep{liang08}. 

In the case of an AWGN channel, the probabilities of detection $P_d$ and false alarm $P_f$ are given as follows \citep{liang08}: 
\begin{eqnarray} \label{eq1}
P_{d,awgn} = Q\left(\left(\frac{\eta}{\sigma_{n}^2}-\gamma -1\right) \sqrt{\frac{\tau f_s}{2\gamma +1}} \right)  
\end{eqnarray}
\begin{eqnarray} \label{eq2}
P_f = Q \left( \frac{\eta}{\sigma_n^2 -1} \sqrt{\tau f_s} \right) 
\end{eqnarray}
where $Q$ is the complementary distribution function of the standard Gaussian:\begin{eqnarray} \label{eq3}
Q(x) = \frac{1}{\sqrt{2 \pi}} \int_x^{\infty} e^{-\frac{t^2}{2}} dt . 
\end{eqnarray} 
In the case of a Rayleigh fading channel, the probability of detection $P_d$ for an average SNR value $\bar{\gamma}$ and based on the chi-square distribution with a normalized parameter $\beta$ as presented in \citep{digham07} is given as follows: 
\begin{eqnarray} \label{eq4}
P_{d,Ray} = e^{-\frac{\eta}{2\sigma^2}} \sum_{i=0}^{N/2-2} \frac{\left(\frac{\eta}{2\sigma^2}\right)^i}{i!} + \left(\frac{2\sigma^2 + \beta \bar{\gamma}}{\beta \bar{\gamma}} \right)^{N/2-1} \nonumber \\
\times \left[ e^{-\frac{\eta}{2\sigma^2 +\beta \bar{\gamma}}} - e^{-\frac{\eta}{2\sigma^2}} \sum_{i=0}^{N/2-2} \frac{\left(\frac{\eta \beta \bar{\gamma}}{2\sigma^2(2\sigma^2 + \beta \bar{\gamma}
)} \right)^i}{i!} \right]
\end{eqnarray} 
where $N$ is equal to the sampling factor $\tau f_s$. 

Fig. 1 depicts the complementary receiver operating characteristics (ROC) curve, which deploys the probability of missed detection $1-P_d$ versus the probability of false alarm $P_f$ for different number of threshold $\eta$, where $\eta \in [0,1]$. Moreover, the sensing time is considered $\tau=2ms$ (solid line) and $\tau=4ms$ (dashed line), while the sensed SNR is considered $\gamma=-15dB$  and  $\gamma=-13dB$ respectively. Obviously, when the sensing time and the sensed SNR increase, the probability of missed detection decreases. Fig. 2 depicts the complementary ROC curve in the case of Rayleigh fading channel for average SNR values equal to $\bar{\gamma}=10dB$ and $\bar{\gamma}=20dB$, while the number of samples are equal to $N=5$ (solid line) and $N=10$ (dashed line). An identical number of samples and sensed SNR are also realized in case of Rayleigh channel.  

\begin{figure}
  \centering
  \includegraphics[width=110mm,height=80mm]{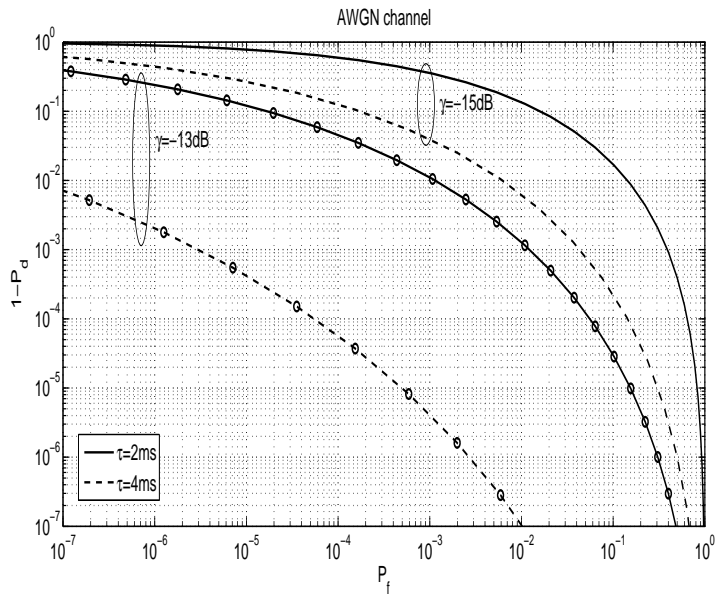}\\
  \caption{Complementary ROC for AWGN channel}
  \label{fig:ROC_awgn}
\end{figure}

\begin{figure}
  \centering
  \includegraphics[width=110mm,height=80mm]{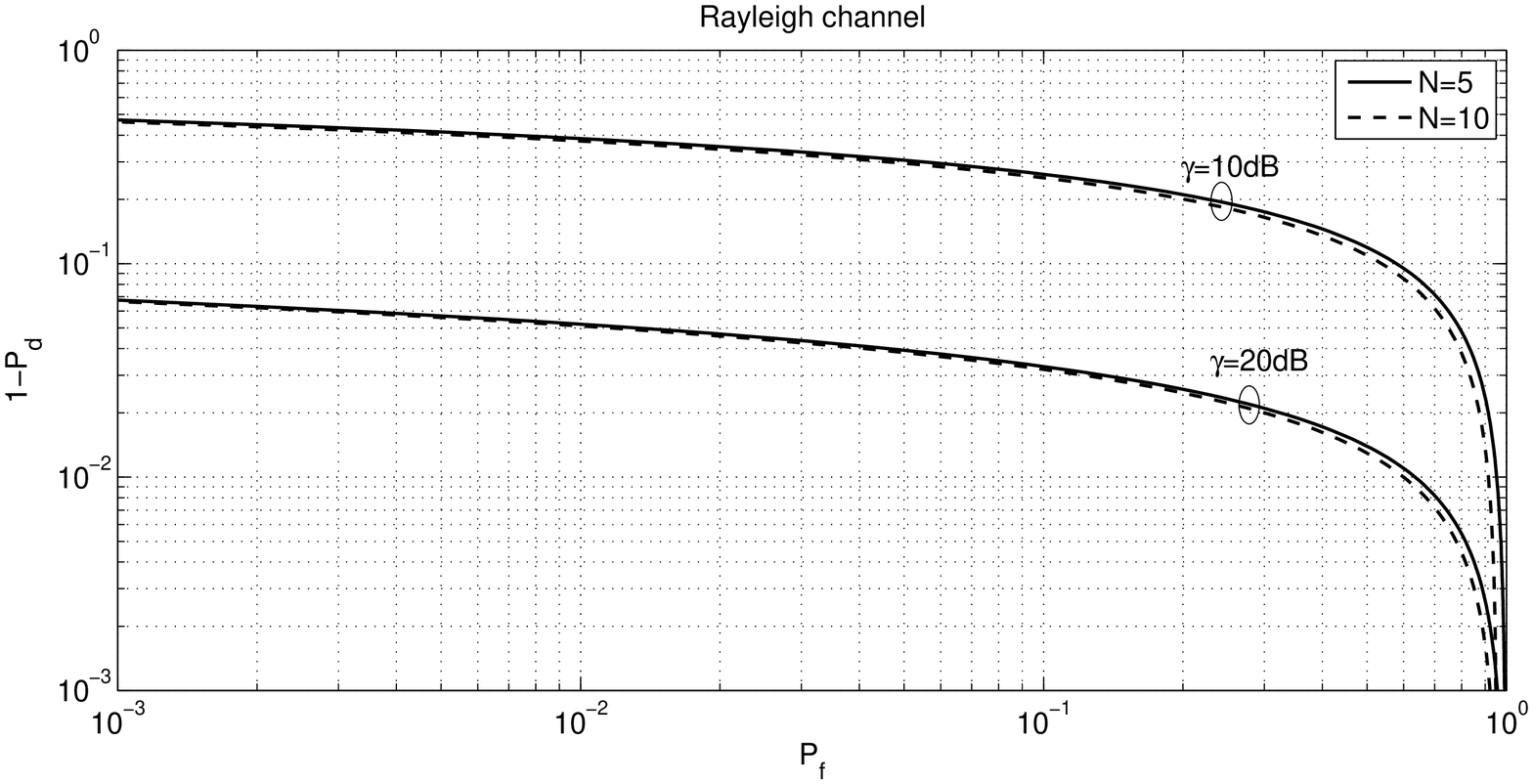}\\
  \caption{Complementary ROC for Rayleigh channel}
  \label{fig:ROC_Ray}
\end{figure}

At the MAC layer, the CSMA/CA protocol for packet transmission is considered in saturation conditions (i.e. each station has always a packet for transmission) with a binary slotted exponential backoff of maximum backoff stage $m$ and minimum contention window of size $W$ \citep{bianchi00}. We assume that for a slot at time $t$, a number of $C$ channels are being sensed and a node of the CRN is able to transmit only in one channel per slot \citep{bianchi00}. Thus, based on \citep{bianchi00} and (Xing, 2006), we model both SpSe and CSMA/CA as discrete time Markov processes denoted as $s_t$ and $b_t$ respectively. In order to cross-layer design (CLD) the SpSe with CSMA/CA, we construct an augmented Markov chain with a state pair $(s_t,b_t)$ containing both the SpSe and the CSMA/CA states (Liu et al., 2005). Fig.1 depicts the concept of the specific CLD where the backoff process $b_t$ at the MAC layer at time $t$ is changed based on the process of SpSe $s_t$ that takes place at the physical layer during the time slot $t$.
\begin{figure}
  \centering
  \includegraphics[width=80mm,height=50mm]{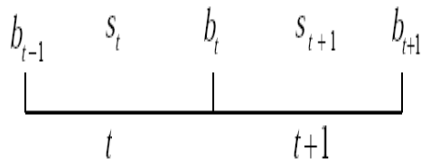}\\
  \caption{CSMA/CA and SpSe as discrete time Markov processes $b_t$ and $s_t$ respectively}
  \label{fig:systemmodel}
\end{figure}

The CSMA/CA at the MAC layer is modelled as a bi-dimensional process ${g(t),b(t)}$ where $g(t)$ is the stochastic process of the backoff stage with size $i\in(0,m)$ and $b(t)$ is the stochastic process of the backoff timer with size $k\in(0,W_{i-1})$ where $W_i$ is equal to $2^{i}W$ \citep{bianchi00}. The SpSe is modelled as a discrete time birth-death process ${s(t)}$ with size $s\in(0,C)$, where one channel per time slot $t$ is considered as active or idle with probability $\alpha$ and $1−\alpha$ respectively (Xing, 2006). All channels are considered homogeneous with equal activity $\alpha$ which highlight's the channel's percentage occupancy by the PU and the probability that a channel is active is $\alpha P_d+(1-\alpha)P_f$ due to the imperfect SpSe. Thus, the CLD of SpSe with CSMA/CA that incorporates the channel’s activity $\alpha$, the spectrum sensing results $P_d$ and $P_f$, the backoff stage $m$ and the minimum contention window $W$ can be modelled as a three-dimensional stochastic process $\lbrace g(t),b(t),s(t)\rbrace$ with $i$, $k$ and $s$ dimensions respectively (Liu et al. 2005). We denote the state transition probability as $P_{\lbrace i,k,s \rbrace \lbrace j,l,u \rbrace}$ that refers to the transition probability from state $\lbrace g_t=i,b_t=k,s_{t-1}=s \rbrace$ , to state $\lbrace g_{t+1}=j,b_{t+1}=l,s_t=u \rbrace$  where $\lbrace i,k,s \rbrace\in[(0,m)\times(0,W_{i-1})\times(0,C)]$.  Fig.2 depicts the derived Markov chain diagram that shows the transition probabilities based on the SpSe results, i.e. idle or busy derived within a discrete time slot $t$. In particular, it presents the changes within a backoff stage as well as between different backoff stages adopting the transitions defined in (Bianchi, 2000) and considering the results of the imperfect SpSe.

\begin{figure}
  \centering
  \includegraphics[width=160mm,height=120mm]{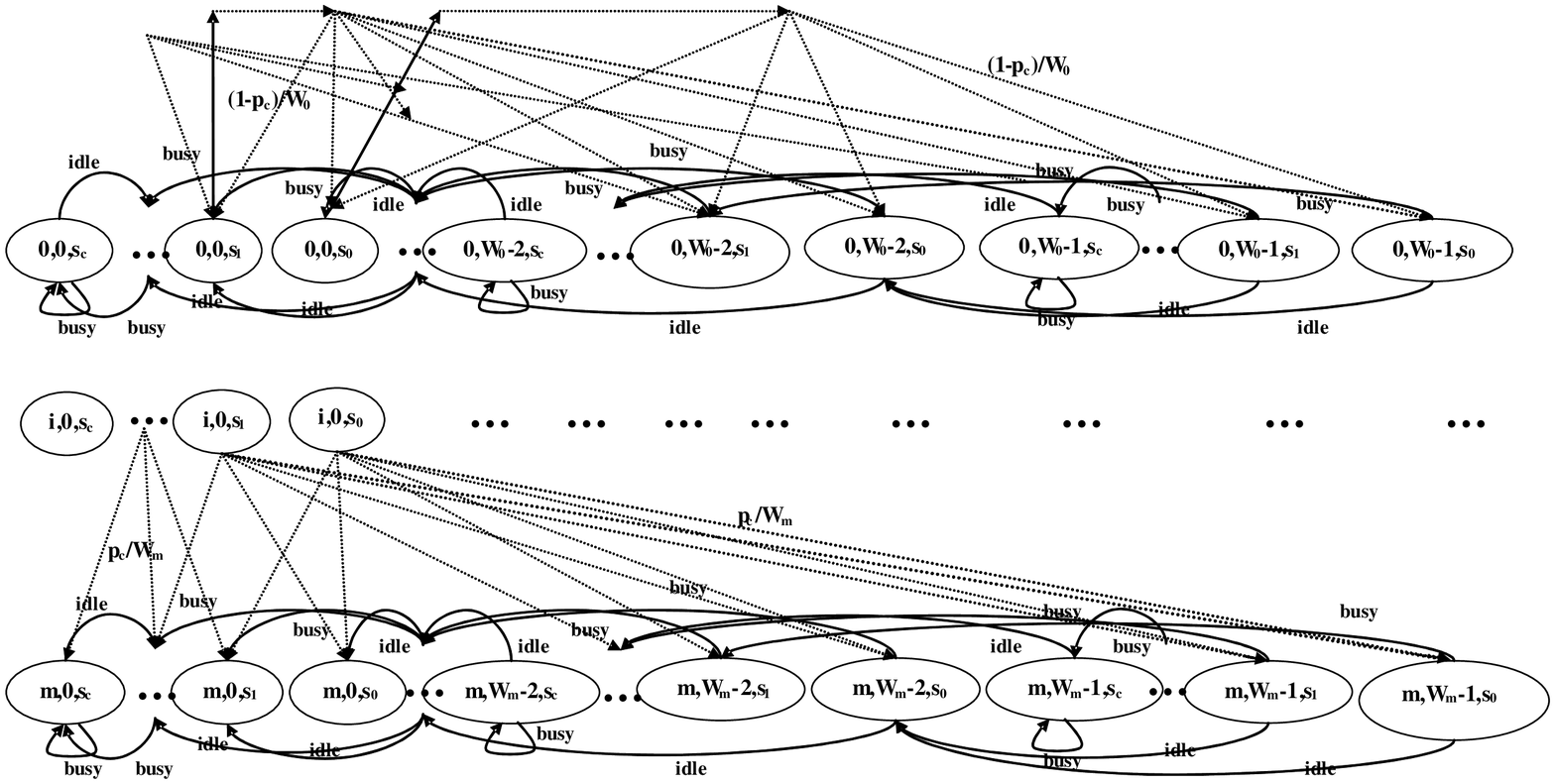}\\
  \caption{The tri-dimensional Markov chain diagram of CSMA/CA with SpSe}
  \label{fig:fig2}
\end{figure}

The state transition probability matrix is organized in a block form where matrix $A$ merges the stochastic process of the CSMA/CA protocol with an exponential backoff with $i\in(0,m)$ size given as follows:
\begin{eqnarray} \label{eq5}
 A =  \begin{bmatrix} \Pi_{\lbrace 0,0\rbrace} & \cdots & \Pi_{\lbrace 0,W_0-1\rbrace} \\ \vdots & \ddots & \vdots \\ \Pi_{\lbrace m,0\rbrace} & \cdots & \Pi_{\lbrace 0,W_m-1\rbrace} \end{bmatrix}
\end{eqnarray}
which includes the sub-matrix $\Pi_{\lbrace i,k \rbrace \lbrace j,l \rbrace}$ that merges the stochastic process of SpSe $\lbrace s(t)\rbrace$ with size $s\in(0,C)$ which is given as follows: 
\begin{eqnarray} \label{eq6}
\Pi_{\lbrace i,k \rbrace \lbrace j,l \rbrace} =  \begin{bmatrix} P_{\lbrace i,k,s_0 \rbrace \lbrace j,l,s_0 \rbrace} & \cdots & P_{\lbrace i,k,s_0 \rbrace \lbrace j,l,s_c \rbrace} \\ \vdots & \ddots & \vdots \\ P_{\lbrace i,k,s_c \rbrace \lbrace j,l,s_0 \rbrace} & \cdots & P_{\lbrace i,k,s_c \rbrace \lbrace j,l,s_c \rbrace} \end{bmatrix} .
\end{eqnarray}

This matrix represents a joint steady state probability that we denote here as $\pi_{\lbrace i,k,s \rbrace}$ that is given as follows: 
\begin{eqnarray} \label{eq7}
 \pi_{\lbrace i,k,s \rbrace} = \lim\limits_{t \rightarrow \infty} P(g_{t-1}=i,b_{t-1}=k,s_t=c) . 
\end{eqnarray}
It is proved that this stationary distribution exists and is unique with the following proposition:

\textit{Proposition:} If $\Pi^t(x,y)$ is the state transition matrix at $t$-step, then for a finite ergodic Markov chain with a finite state space $\Omega$, there exists a unique stationary distribution $\pi$, such that:
\begin{eqnarray} \label{eq8}
\lim_{t \rightarrow \infty} \Pi^t(x,y) = \pi(y), & & \forall (x,y) \in \Omega
\end{eqnarray}

\textit{Proof:}  For all $x,y \in \Omega$, a distribution $\pi$ is a stationary distribution, if it is invariant with respect to the transition matrix, which is expressed as follows:
\begin{eqnarray} \label{eq9}
\pi(y) = \sum_{x \in \Omega}^m \pi(x) \Pi(x,y), & & \forall (x,y) \in \Omega
\end{eqnarray}
It holds that:
\begin{eqnarray} \label{eq10}
\sum_{x \in \Omega} \pi(x) \Pi(x,y) &=& \sum_{x} \left( \lim_{t\rightarrow \infty} \Pi^t(z,x) \right) \Pi(x,y),  \forall z \in \infty \nonumber \\
&=& \lim_{t \rightarrow \infty} \left( \sum_{x} \Pi^{t}(z,x) \Pi(x,y) \right) \nonumber \\
&=& \lim_{t \rightarrow \infty} \left( \sum_{x} \Pi^{t+1}(z,y)  \right)  \nonumber \\
&=& \lim_{t \rightarrow \infty} \left( \sum_{x} \Pi^{t}(z,y) \right) \nonumber \\
&=& \pi(y)
\end{eqnarray}
which shows that the $\pi$ is an invariant distribution and thus a unique stationary distribution. 

The stationary distribution $\pi$ is a row vector defined as follows: 
\begin{eqnarray} \label{eq11}
 \pi = [\pi_{0,0,0},\cdots,\pi_{0,W_{i-1},0},\cdots,\pi_{0,W_{i-1},C},\cdots, \nonumber \\
 \pi_{m,0,0},\cdots,\pi_{m,W_{i-1},0},\cdots,\pi_{m,W_{i-1},C}]
\end{eqnarray}
for which the following holds:
\begin{eqnarray} \label{eq12}
\sum_{i,k,s} \pi_{\lbrace i,k,s \rbrace} = \sum_{i=0}^m \sum_{k=0}^{W_i-1} \sum_{s=0}^C \pi_{\lbrace i,k,s \rbrace}
\end{eqnarray}
Based on this joint stationary distribution, we can derive the transmission and collision probabilities and the achievable throughput as we present below.

\section{Performance Metrics}

\subsection{Collision Probability}

In the proposed scheme, a transmission occurs when the backoff timer is equal to zero i.e. $k=0$ and when at least one channel is idle i.e. $s\in(0,C-1)$. Thus, the transmission probability is defined as follows: 
\begin{eqnarray} \label{eq13}
 \tau = \sum_{i=0}^m \sum_{s=0}^{C-1} \pi_{\lbrace i,0,s \rbrace}
\end{eqnarray}
Since the SpSe and the CSMA/CA are mutually independent, the transmission
probability for a specific backoff stage $i\in(0,m)$, backoff timer $k\in(0,W_{i-1})$ and spectrum sensing result $s\in(0,C)$ can be expressed as: 
\begin{eqnarray} \label{eq14}
 \tau(i,k,s) = \sum_{i=0}^m b_{i,0} \sum_{s=0}^{C-1} s_c
\end{eqnarray}
where $b_{i,0}$ and $s_c$ are the stationary probabilities of the CSMA/CA and the SpSe processes respectively (Bianchi, 2000) (Xing et al. 2006). The collision probability $p_c$ that begets at the SN when at least one of the $n - 1$ remaining stations of the PN transmits on the same channel is derived as follows [Chong et al., eq.(9), 2010]: 
\begin{eqnarray} \label{eq15}
 p_c(i,k,s) = 1-(1-\tau)^{n-1}
\end{eqnarray}

\subsection{Normalized Throughput}

We analyze now the throughput that is achieved from the collision avoidance protocol using the considered CLD. The advantage of this approach is that both transmission and collision probabilities in equations \eqref{eq14} and \eqref{eq15} incorporate the parameters of CSMA/CA and SpSe mechanisms and thus a normalized throughput can incorporate the expected CLD results as well. In general, the throughput is defined as the information transmitted in a slot time divided with the total slot duration (Bianchi, 2000). We will follow the definition of the normalized throughput since the aim of this work is to highlight the functionality of collision avoidance protocols in CRNs. In this context, (Chong et. al, 2010) derived an identical formula; however, our formulation is derived using the proposed CLD and thus it merges all the involved parameters in the simulation results. Subsequently, the normalized throughput is defined as follows: 

\begin{eqnarray} \label{eq16}
 S = \frac{\rho P_{Tr}}{\rho(P_{Tr})+P_{Fr}+P_{C}}
\end{eqnarray}
where the $P_{Tr}$, $P_{Fr}$ and $P_C$ are the probabilities that there is a successful transmission in the considered slot with duration $\rho$, the channels are not occupied and the transmission may face collisions that are defined as follows: 
\begin{eqnarray} \label{eq17}
 P_{Tr} = n \tau (1-\tau)^{n-1}
\end{eqnarray}
\begin{eqnarray} \label{eq18}
 P_{Fr} = (1-\tau)^n
\end{eqnarray}
\begin{eqnarray} \label{eq19}
 P_C = 1-P_{Tr}-P_{Fr}
\end{eqnarray}
By substituting equations \eqref{eq17}, \eqref{eq18} and \eqref{eq19} into equation \eqref{eq16}, the normalized throughput according to the considered CLD is given as: 
\begin{eqnarray} \label{eq20}
 & & S(i,k,s) = n \tau (1-\tau)^{n-1} \nonumber \\ 
& &  \forall i\in (0,m),\forall k\in (0,W_{i-1}),\forall s\in (0,C) 
\end{eqnarray}
Based on these metrics and by varying the corresponding parameters, we assess the performance of the CSMA/CA protocol in TV white bands. 

\section{Results}

We have derived both numerical and simulation results in order to validate the proposed CLD using the simulation method for Markov chains provided in \citep{tranter03}. We assume that the sensed SNR is equal to $\gamma=-15dB$ and the sensing time equal to $T_s=2ms$ for primary channels with frequency $f_p=6MHz$ that can be considered as the optimum value in the case of IEEE 802.22 as indicated by \citep{liang08}. Notably, we use a fixed probability of detection $P_d$ in order to get the results, which gives a fixed probability of false alarm $P_f$ using the eq.(13) in \citep{liang08} \footnote{$P_f=Q(\sqrt{2\gamma +1}Q^{-1}(P_d)+\sqrt{\tau f_s }\gamma)$}. An alternative way of calculating the SpSe parameters is to change the sensing threshold $\eta$ and time $\tau$ separately; however, such an option requires an optimization technique over several PHY and MAC parameters, which is actually out of the scope of this paper. The reader can refer to \citep{liang08} for information related to optimization on sensing time $\tau$ and to \citep{foukalas12} for optimization on the sensing threshold $\eta$.  

Fig.5 depicts the collision probability $p_c$ versus the number of stations $n$ for different probabilities of detection $P_d$ equal to $1,0.9,0.1$ and minimum contention window $W$ for a constant backoff stage $m=3$. The results are  taken considering one channel i.e. $C=1$ assuming an activity probability equal to $\alpha=0.5$. The solid lines depict the case of $m=3$ and $W=32$ and the dashed lines depict the case of $m=3$ and $W=64$. The simulation results are depicted without lines and they demonstrate that the analytical model is accurate since its results almost coincide with the simulation ones. It is obvious that a low probability of detection e.g. $P_d=0.1$, results in high collision probability $p_c$. On the other hand, a high contention window value, e.g. $W=64$ (dashed lines), results in a lower collision probability $p_c$, since a high contention window value results in higher transmission probability. Fig. 6 depicts again the collision probability $p_c$ versus the number of stations $n$ for different probabilities of detection $P_d$ with a backoff stage equal to $m=5$ for a constant minimum contention window $W=32$. It also depicts the simulation results without lines. It can be observed that the changes in the backoff stage induce less decrease in the collision probability $p_c$ than that induced by the changes in the contention window.

\begin{figure}
  \centering
  \includegraphics[width=120mm,height=90mm]{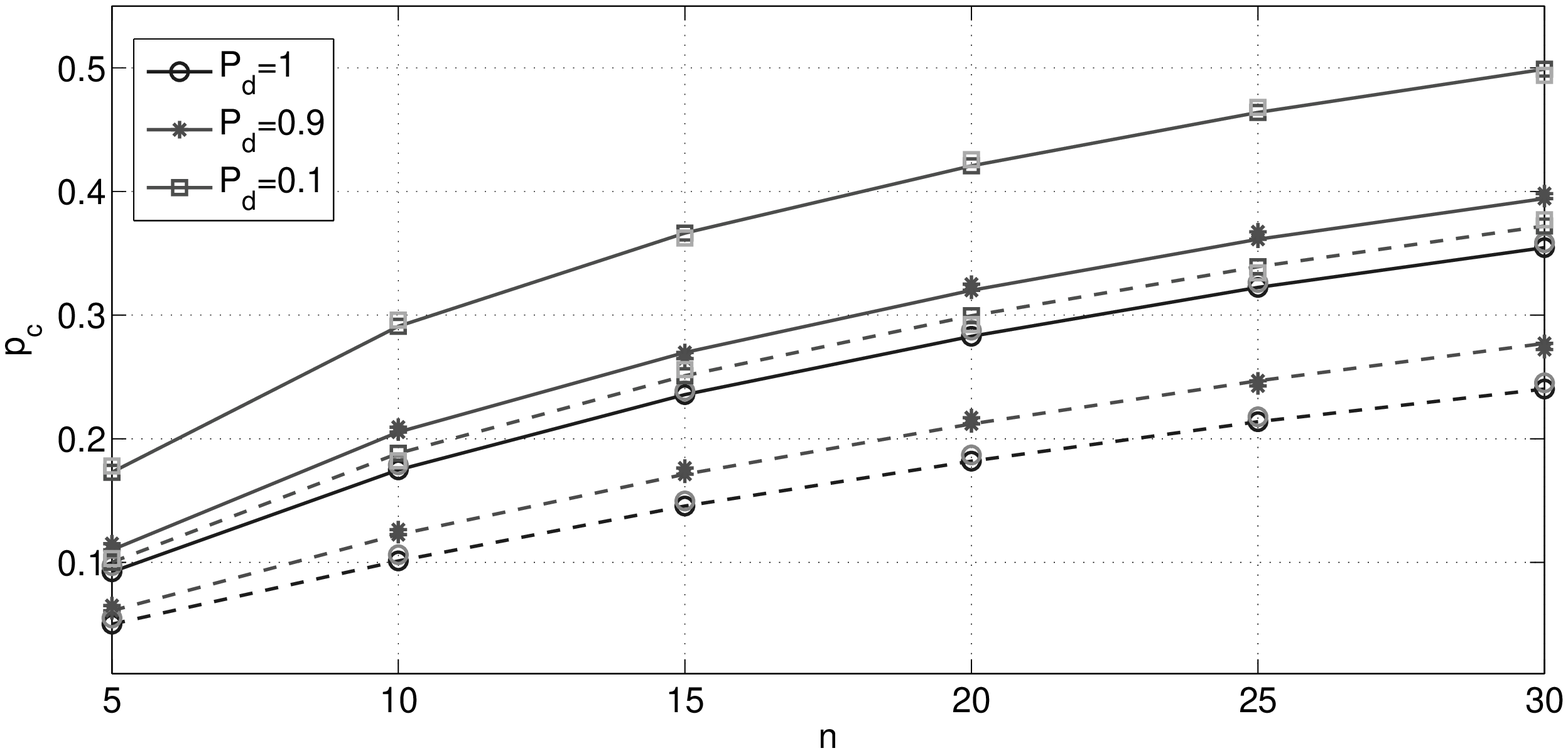}\\
  \caption{Collision probability $p_c$ vs. number of stations $n$ for different probabilities of detection $P_d$ with backoff stage $m=3$ and contention window $W=32$ (solid lines) and with $m=3$ and $W=64$ (dashed lines)}
  \label{fig:3}
\end{figure}

\begin{figure}
  \centering
  \includegraphics[width=120mm,height=90mm]{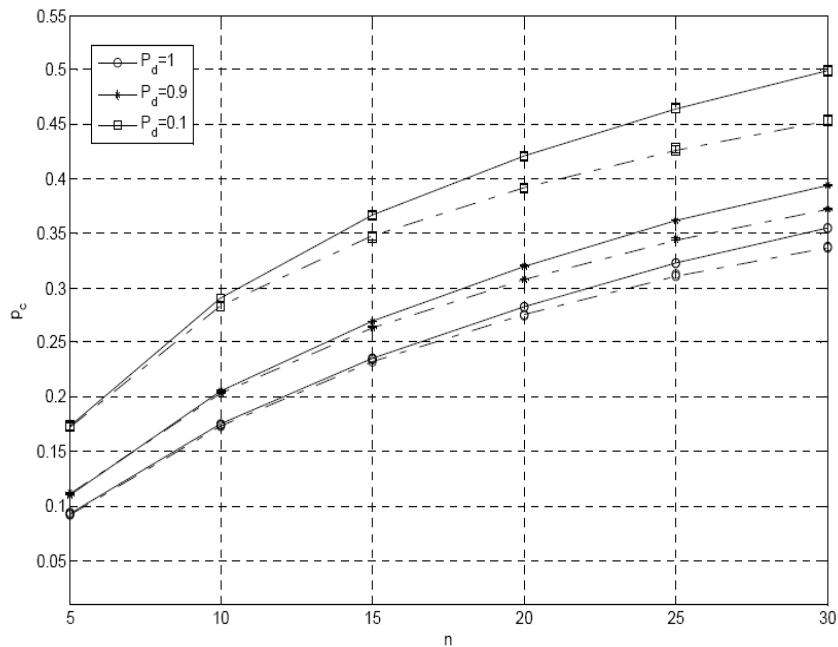}\\
  \caption{Collision probability $p_c$ vs. number of stations $n$ for different probabilities of detection $P_d$ with backoff stage $m=3$ and contention window $W=32$ (solid lines) and with $m=5$ and $W=32$ (dashed lines)}
  \label{fig:4}
\end{figure}

Fig.7 depicts the collision probability $p_c$ versus the number of stations $n$ for different values of channels’ activity $\alpha$ and number of channels $C$. The results are obtained considering a probability of detection equal to $P_d=0.5$, a backoff stage equal to $m=3$ and a contention window equal to $W=32$. The solid lines depict the case of $\alpha=0$, $\alpha=0.5$, $\alpha=0.8$ and $C=1$, the dashed lines depict the case of $C=3$ and the dotted dashed lines the case of $C=6$ considering the same activities for all cases. We also depict the simulation results without lines. Obviously, a high probability of activity (e.g. $\alpha=0.8$) results in a lower collision probability $p_c$. Furthermore, for a high number of sensed channels e.g. $C=6$, the collision probability $p_c$ increases.

\begin{figure}
  \centering
  \includegraphics[width=120mm,height=90mm]{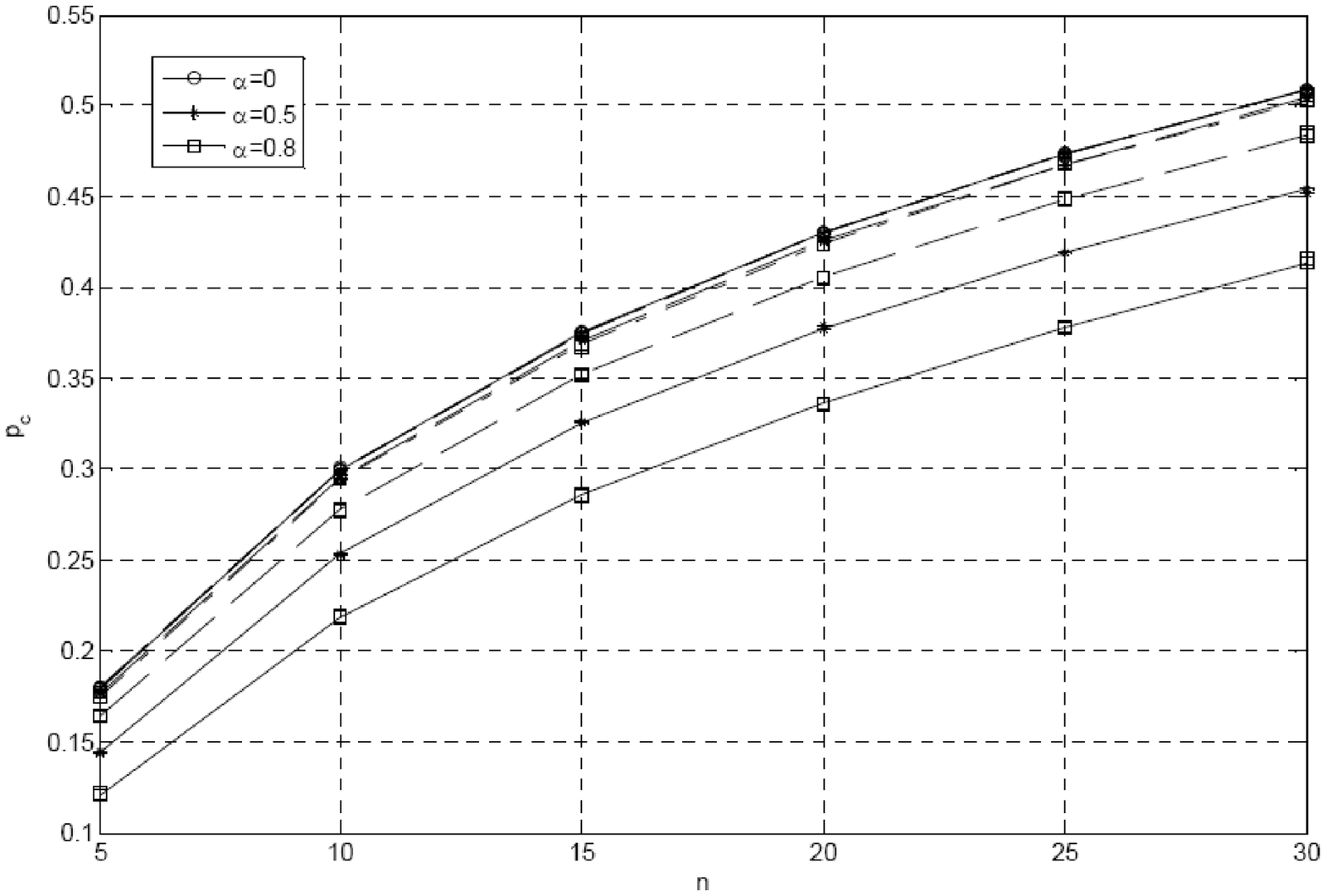}\\
  \caption{Collision probability $p_c$ vs. number of stations $n$ for different activities $\alpha$ with $C=1$ channels (solid lines), with $C=3$ (dashed lines) and with C=6 channels (dotted dashed lines)}
  \label{fig:5}
\end{figure}

Fig. 8, Fig. 9 and Fig. 10 depict the normalized throughput $S$ versus the probability of detection $P_d$ by varying contention window size $W$ and backoff stage $m$ and the number of channels upon, which the SpSe mechanism is accomplished. Notably, simulation results are not depicted in these figures since their correctness is verified from the respective collision probability results. A general outcome could be expressed as the fact that an increase in the probability of detection $P_d$ decreases the normalized throughput $S$ and this is expected due to the decrease in the transmission probability $\tau$. On the contrary, when the number of stations $n$ increases the transmission probability $\tau$ increases as well. 

In particular, Fig.8 depicts the results for contention window size $W=32$ and $W=64$ and for different number of stations n considering a PU’s activity equal to $\alpha=0.5$ for $C=1$ number of channels and backoff stage equal to $m=3$. Obviously, a decrease of the throughput due to an increase on the contention window is observed, i.e. $W=64$ has an impact on the performance even in the case of a high number of stations, e.g. $n=10$. However, an identical outcome does not appear in case that the backoff stage increases as shown in Fig.9, which depicts the normalized throughput $S$ versus the probability of detection $P_d$ for $m=3$ and $m=5$ backoff stages considering a PU’s activity equal to $\alpha=0.5$ for $C=1$ number of channels and contention window equal to W = 32 . Obviously, a decrease of the throughput due to an increase on the backoff stages, i.e. $m=5$ is not so high when the number of stations increases, e.g. $n=10$. On the other hand, for a small number of stations, the decrease of the throughput due to the increase of backoff stage is negligible. 
\begin{figure}
  \centering
  \includegraphics[width=120mm,height=90mm]{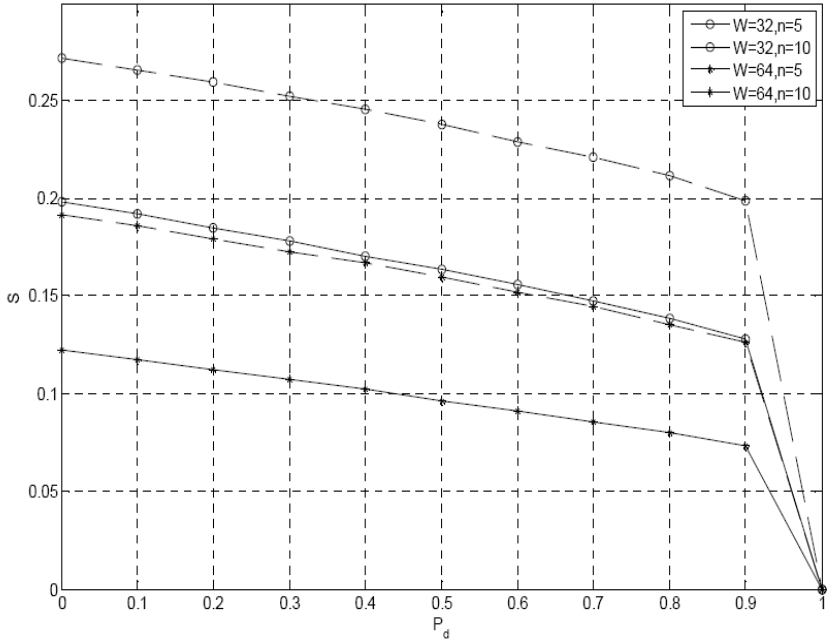}\\
  \caption{Normalized throughput $S$ vs. probability of detection $P_d$ for different number of stations $n$, for PU’s activity $a=0.5$ for $C=1$ and backoff stage $m=3$ with contention window $W=32$ and $W=64$}
  \label{fig:6}
\end{figure}
\begin{figure}
  \centering
  \includegraphics[width=120mm,height=90mm]{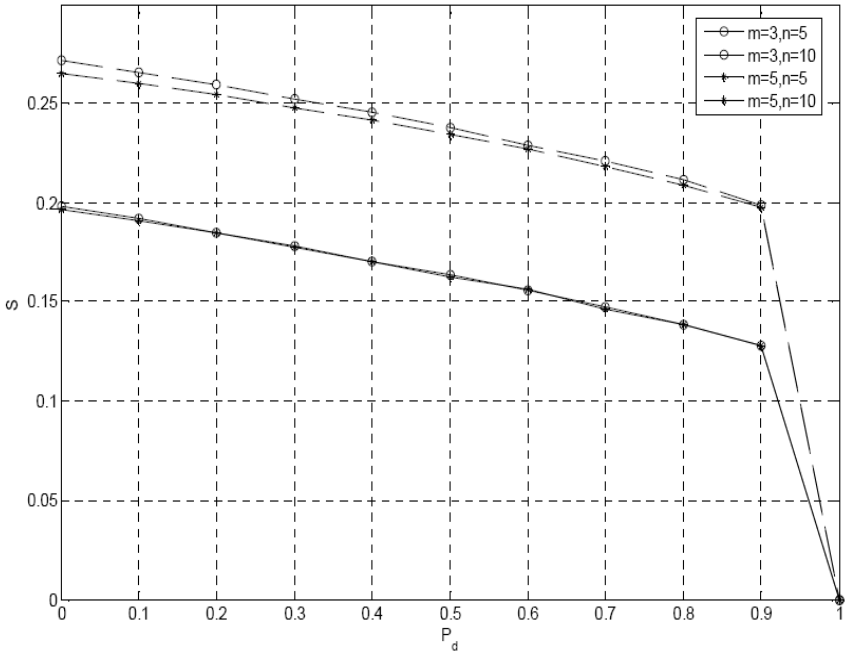}\\
  \caption{Normalized throughput $S$ vs. probability of detection $P_d$ for different number of stations $n$, for PU’s activity $a=0.5$ for $C=1$ and contention windows $W=32$ with backoff stage $m=3$ and $m=5$}
  \label{fig:7}
\end{figure}

Finally, Fig.10 depicts the normalized throughput $S$ versus the probability of detection $P_d$ for different number of channels $C$ considering a PU’s activity equal to $\alpha=0.5$, backoff stage equal to $m=3$ and contention window equal to $W= 32$. Obviously, changing the number of sensed channels, the throughput increases. However as the number of channels being sensed increases the throughput performance is not as high as in the cases of $C=3$ and $C=6$ channels.
\begin{figure}
  \centering
  \includegraphics[width=120mm,height=90mm]{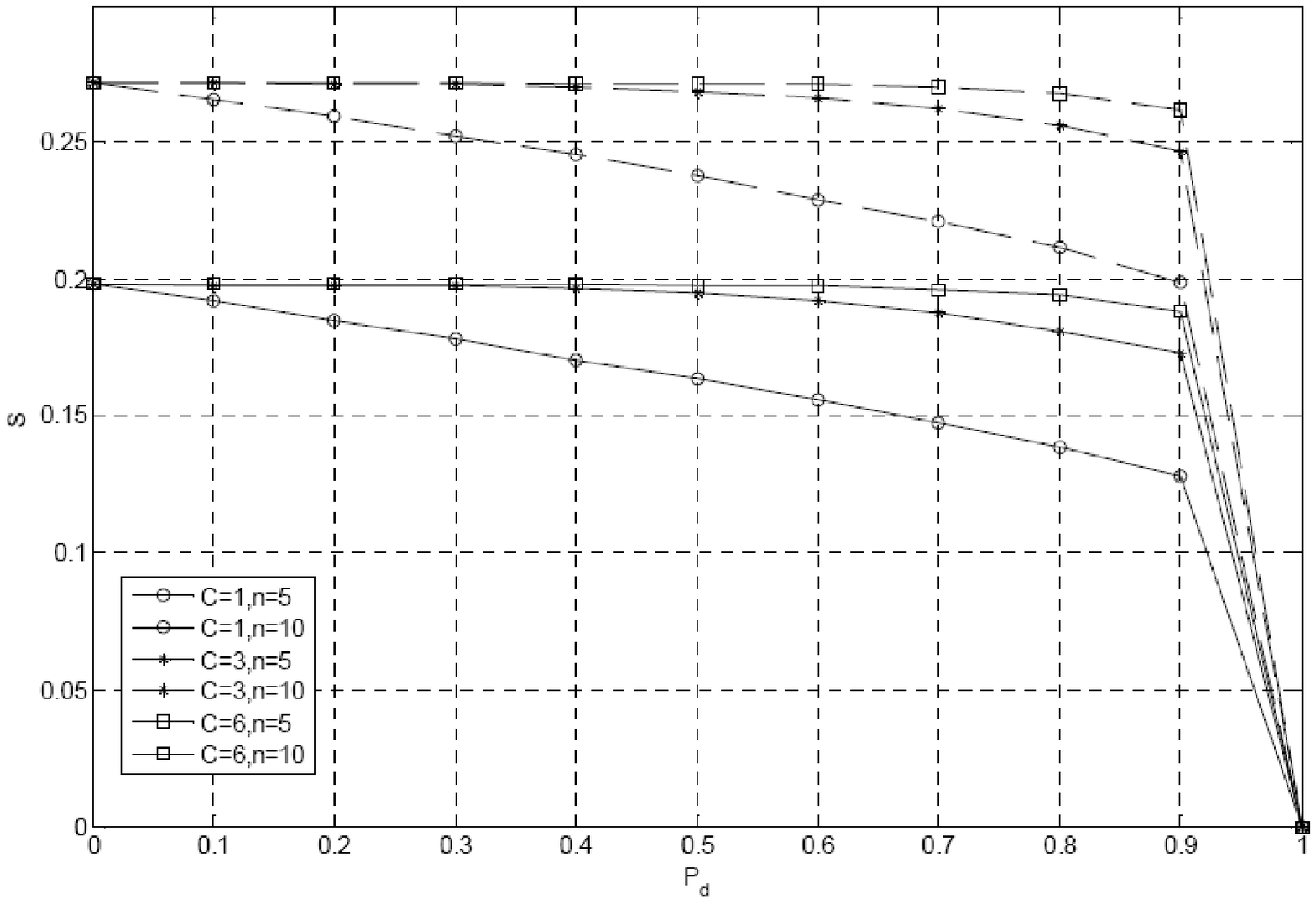}\\
  \caption{Normalized throughput $S$ vs. probability of detection $P_d$ for different number of stations $n$, for PU’s activity $a=0.5$, backoff stage $m=3$ and contention windows $W=32$ and for different number of channels $C$}
  \label{fig:8}
\end{figure}

\section{Conclusions}
In this paper, we introduced and investigated a cross-layer design of the spectrum sensing mechanism and CSMA/CA protocol at the PHY and MAC layers respectively. Using such a design, we derived the collision probability and the corresponding achievable throughput in CRNs. We relied on a discrete time Markov chain model with a state pair for modelling both spectrum sensing mechanism and the CSMA/CA protocol. The considered model resulted in a joint stationary probability which incorporated the parameters of spectrum sensing and CSMA/CA, which were used in the sequel for the derivation of the collision probability and normalized throughput. Simulation and numerical results are provided, which reveal the performance of the proposed CLD, that can be deployed practically within a DCF in CRNs. The corresponding outcomes illustrate the performance that can be achieved and the gain that can be anticipated from the application of the proposed approach for enhancing the utilization of TVWSs that may be idle in time and space.

\section*{Acknowledgement}
This research has been co-financed by the European Union (European Social Fund – ESF) and Greek national funds through the Operational Program "Education and Lifelong Learning" of the National Strategic Reference Framework (NSRF) - Research Funding Program: ARCHIMEDES III. Investing in knowledge society through the European Social Fund.

\end{document}